# AI Agents Executing an RMS and EMT Ride-Through Workflow on a 2,000-Bus Synthetic Texas Grid: A Case Study with Open-Source Tools

Tam Kemabonta Brighton Ombuki Vladimir Abdelnour

*Abstract*—**This paper reports a case study of engineer-supervised AI agents executing a predefined RMS and EMT ride-through workflow on the 2,000-bus ACTIVSg2000 synthetic Texas grid with open-source tools. The test case is a 133 MW battery with a co-located 130 MW data-center load, assessed against ride-through criteria fixed before simulation and for loss of synchronism elsewhere in the grid. One agent ran 44 RMS cases and 15 without-project comparisons in ANDES; the other ran a 220-row EMT intake bench and an 84-row system study in DPsim and ParaEMT on a Kron-reduced area. Across the 44 RMS cases, 127 criterion cells fail, including loss of synchronism in all 44. Two findings concern the project plant. The battery keeps absorbing reactive power during sags on the light-load schedule (B3, as low as −19.8 Mvar on S4 T1), and the data-center load returns only at 5.00–5.56 s after long bolted faults (L3), which misses the ERCOT load-return requirement. In 13 of the 15 comparisons, which the authors selected after slips or speed-gate stops, the same machines slip without the project; the other 29 runs have no without-project comparison. RMS and EMT values differ in network, plant model, controls, operating point, and base, and so they are reported as separate, unaligned observations. The paper reports the errors the authors identified in the agent logs, 63 author messages, and 3.8 days from agent start to the integrated study report. The agents executed and documented the predefined ride-through and stability workflow, which covers one part of an interconnection study.**

*Keywords*— **AI agents, electromagnetic transients, interconnection studies, large language models, transient stability**

## I. INTRODUCTION

At the end of 2025, about 2,061 GW of generation and storage capacity was actively seeking transmission interconnection in the United States, 1,312 GW of generation and 749 GW of storage [1]. Root-mean-square (RMS) stability studies, electromagnetic transient (EMT) model intake tests, and EMT system studies assess complementary aspects of inverter-based resource (IBR) performance [2], [3], and EMT studies assess behavior that positive-sequence models may not capture [2]. ERCOT filings use PSS/E, a positive-sequence package, and PSCAD, an EMT package [3]. Large computational loads add ride-through obligations: Section 2.15 of the ERCOT Nodal Operating Guides sets voltage ride-through requirements for them [4], and a May 2026 NERC guideline recommends that large-load entities maintain site-specific EMT models, which a transmission planner may require [5]. Software agents built on large language models now attempt such work.

Earlier agent work drives power-system simulators: a feedback-driven multi-agent framework for simulations [6], generation of EMT models of the IEEE 9-bus system [7], and generation of ANDES disturbance scenarios for transient stability on the IEEE 39- and 118-bus systems [8].

The test case is a 133 MW battery with a co-located 130 MW data-center load, connected at a 230 kV bus of a public 2,000-bus synthetic grid and assessed against ride-through criteria fixed before simulation and for loss of synchronism elsewhere in the system. The study question is what engineer-supervised AI agents can execute in a predefined RMS and EMT workflow that addresses it, at what cost in time and supervision, and with which errors. Its scope is the ride-through and stability part of an interconnection study, with each simulator's results reported on their own terms. The contributions are:

- an RMS and EMT ride-through workflow with criteria hashed before any fault, executed by two AI agents;
- a record of the errors identified in the agent logs, how each was detected, the 63 author messages, and the elapsed time;
- a statement of how the tool setups differ, what the cross-tool observations show, and what a validated comparison would require.

## II. STUDY SYSTEM AND WORKFLOW

### A. Network, Project, and Criteria

The network is the ACTIVSg2000 synthetic Texas grid [9], [10]. The project is a 133 MW battery energy storage system

(BESS) and a 130 MW data-center load (75% information technology on an uninterruptible supply, 25% cooling, 0.98 power factor) behind one point of interconnection (POI), bus 6255 at 230 kV, through a main transformer. The positive-sequence short-circuit level at the POI is 17.20 GVA with X/R of 9.16 (100 MVA base), computed at bus 6255 from the full 2,000-bus admittance matrix with synchronous machines behind subtransient reactance, constant-impedance loads at the power-flow voltage, and IBRs excluded. The four dispatches are summer peak (S1, 67.1 GW), light load with the battery charging (S2), summer peak with a 500 MW data center (S3), and S2 with the project at a 115 kV weak tap, bus 6018 (S4).

Eleven RMS contingencies (T1–T11) run on each dispatch, giving 44 RMS cases: three-phase faults near the POI and on a 500 kV line, a single-line-to-ground (SLG) fault, a stuck breaker, six sags, moderate sags, a 2,587 MW generation trip (T8), a data-center load trip, a 1,500 MW load loss, and an overvoltage. Each case is scored against 15 criteria for the battery (B), the load (L), and the system (Y), summarized in Table I (PG: Planning Guide). The B criteria follow PRC-029-1 and IEEE Std 2800-2022 [11], [12], the L criteria follow NOG Section 2.15 [4], and the Y criteria follow ERCOT planning practice [13] and, for synchronous units, PRC-024-4 [14]. The criteria and tolerances were written to a file, hashed with SHA-256 (a cryptographic hash that detects any later change), and archived before any fault was run.

TABLE I. LOCKED CRITERIA (COMPACT FORM)

| ID | Requirement | Source |
|---|---|---|
| B1 | No trip or momentary cessation inside the voltage–time envelope (for example 0.50–0.70 pu for 3.0 s) | NOG 2.9.1.1; PRC-029 |
| B2 | Active power back to the pre-disturbance level within 1.0 s of voltage between 0.90 and 1.10 pu | NOG 2.9.1.1; PRC-029 |
| B3 | Reactive current injected through the mandatory region | PRC-029-1 |
| B4 | No trip within the frequency bands and a rate of change of frequency of 5 Hz/s | NOG 2.6.2.1 |
| L1–L3 | Load stays connected in the envelope; returns to 90% of pre-disturbance power within 2 s of voltage above 0.90 pu | NOG 2.15 |
| L4–L6 | Current at most 150% of normal for 0.5 s; survives six sags in 90 s; stays connected for 58.8–61.2 Hz | NOG 2.15, 2.6.4 |
| Y1 | Buses above 100 kV recover to at least 0.90 pu within 5 s of clearing (10 s for T4, T6) | PG 4.1.1 |
| Y2 | Oscillations of 0.2–2 Hz damped at 3% or more (Prony fit) | PG 4.1.1 |
| Y3 | No loss of synchronism (rule in Section III-C) | PRC-024-4 |
| Y4, Y5 | T8 nadir above 59.3 Hz; load-loss peak below 60.6 Hz and 230/500 kV voltage at most 1.05 pu after 10 s | NOG 2.6.1; PG 4.1.1 |

## B. Workflow and Author Oversight

Two AI agents performed the work, which the authors scoped. The RMS agent ran ANDES on the full network, and the EMT agent ran DPsim and ParaEMT on a reduced area. Both ran as Cursor cloud agents [15] on separate virtual machines using the Grok 4.7 model (grok-4.7-high-fast). Each agent received a written instruction file approved by an author: record hashes of public data, lock the criteria before any fault, never loosen a criterion or tolerance, log every attempt, and stop and report when blocked. The agents exchanged results through versioned archives with SHA-256 manifests, and the EMT agent cross-checked the RMS boundary export against its own network. The authors reviewed reports and sent check-in messages, which kept the RMS runs open because a virtual machine computed only while a run was open. Each numerical change to the RMS solver had to be shown against reference traces before use.

# III. TOOLS AND MODELS

## A. Simulators and Plant Models

ANDES 2.0.0 [16] ran the full network with the trapezoidal method at 1/240 s, a KLU sparse solver, and a Newton tolerance of $10^{-4}$. DPsim 1.4 [17] and ParaEMT [18] ran three-phase EMT, a nodal solution with trapezoidal companion models. Each tool used its own network, plant model, controls, operating point, and base, and the EMT plant is an averaged grid-following model adapted from an open NREL design [19].

TABLE II. TOOL AND MODEL DIFFERENCES AND WALL-CLOCK OBSERVATIONS

| Item | ANDES | DPsim | ParaEMT |
|---|---|---|---|
| BESS, POI transformer | 133 MW; 133 MVA, 8%, X/R 40 | Same | Same |
| Domain | RMS, positive sequence | EMT, three phase | EMT, three phase |
| Network | Full, 2,000 buses | Kron-reduced, 21 buses (S4: 7) | Same reduced (6 cases) |
| Step | 1/240 s | 10 and 20 µs | 50 µs |
| Plant model | REGCA1, REECA1, REPCA1 | NREL-based GFL plant, LCL filter | Same plant, in driver |
| Controls | 1.1 pu limit, reactive priority | PLL, k-factor Q path, 1.1 pu limit | Same as DPsim |
| Fault | 0.001 pu shunt; SLG as positive-sequence shunt | 0.001 Ω or SLG resistance; clearing ramp | Same schedule |
| Plant base | 133 MVA | 140 MVA | 140 MVA |
| Pre-fault Q (S1) | 26.4 Mvar | 1.6 Mvar | 1.4 Mvar |
| Wall time, S1 | 365–366 s for 20 s (18 s/s) | 244–246 s at 20 µs (54 s/s), 490–498 s at 10 µs (110 s/s), for 4.5 s | 38–42 s for 4.5 s (8–9 s/s) |

EMT: zero reactive order, Q at the 24.9 kV bus; plant configuration GTX-GFL-NREL-2021-ERCOT-11-01-2026-v1.2. ParaEMT: S1 E1, E2, E5, E8, S2 E2, S4 E1. The resistance-ramp breaker model of Section V, whose before-and-after pairs are from an earlier plant configuration (DPsim, 50 µs), was carried into this study.

The last row of Table II gives wall-clock observations as the ratio ρ of wall-clock to simulated seconds. Because network scopes and time steps differ, each ratio describes that tool's own runs. The RMS virtual machine had 4 cores and 15 GB of memory with no swap; three workers were pinned to cores 0–2 with BLAS threads set to one, and each timing run used one core. The EMT machine configuration was not recorded.

### B. *Kron Reduction of the EMT Area*

The EMT area keeps every bus within two branches of the POI at or above 115 kV, plus the three Austin 2 gas turbines (buses 6257, 6258, 6260) as explicit machines. The rest of the grid is eliminated by Kron reduction [20],

$$Y_{\mathrm{eq}} = Y_{bb} - Y_{bi}\, Y_{ii}^{-1}\, Y_{ib} \tag{1}$$

where $b$ and $i$ index boundary and internal buses, and each boundary bus is an ideal voltage behind the reduced R–L branch. For S1–S3 the reduced network has 21 buses and 76 branches, with 14 boundary buses and 7 internal buses, including the POI bus 6255. For S4 it has 7 buses and 18 branches, with boundary buses 6175, 6233, 6255, and 6330. Passivation dropped 45 mutual terms below 1%. Reconstruction errors were $1.327\times10^{-9}$ pu and $1.882\times10^{-8}$ degrees. In the S1 summer-peak EMT build, the short-circuit level at the POI differed by 0.1% (X/R of 9.03 and 9.05) between the full network and the passivated reduced network. The check compares the reduced network with the same build's full-network value (17.585 GVA), which was computed separately from the 17.20 GVA reference of Section II-A. The ANDES power flow and the EMT agent's own power flow gave pre-fault bus voltages within 0.005 pu and 0.5° on S1 at every compared bus except bus 6229 (1.0184 and 1.0356 pu); on S2, the differences reach 0.048 pu and 11.2°. This comparison is between two power-flow solutions; the reduced-to-full network check is the short-circuit comparison above.

### C. *Loss-of-Synchronism Rule*

An ANDES run ends early when a machine speed stays at or above 1.20 pu or at or below 0.80 pu for 0.05 s. After the run, a machine is declared out of step when its unwrapped angle relative to the inertia-weighted center of inertia (COI) of the machines that stay in step (the set $S$) [21],

$$\delta_i^{\mathrm{COI}}(t) = \delta_i(t) - \frac{\sum_{j\in S} H_j\, \delta_j(t)}{\sum_{j\in S} H_j} \tag{2}$$

exceeds 360° or is still diverging at the end of the record. The reported loss time is the first sample past 180° [21]. Six runs still opening at the end of the record (S2 T1–T4 and T10, and S4 T10) are scored as past 180° at the last sample, with speed at least 3% from nominal and still moving the same way, so their effective threshold is 180°. Under this rule 38 runs exceed 360° and 6 are still opening, so Y3 fails in 44 runs. Five files were ended by an earlier stop rule and not rerun, each already past 180°: S1 T6, S3 T6, and S2 T1 with the project (38.33, 37.60, and 1.41 s) and S1 T10 with and without it (46.9 and 46.8 s). The locked criterion text states a COI angle separation below 180° and was not edited. Only the stop rule was replaced, on Nov. 1, 2026, because the earlier stop (180° with a 3% speed deviation) produced false stops.

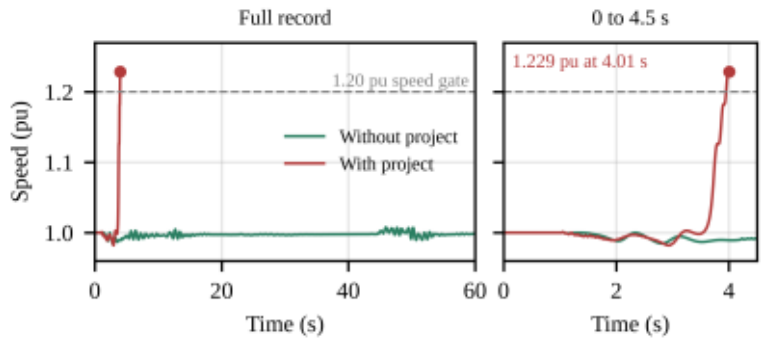

Fig. 1. Buchanan unit 1 speed, S4 T8 (ANDES). With the project, the run ends at the speed gate at 4.01 s; without it, the unit stays in step for 60 s.

## IV. RESULTS

### A. *RMS Cases and Without-Project Comparisons*

Across the 44 RMS runs, 127 criterion cells fail, and Y3 fails on every run. Two plant findings concern the project: on the light-load charging schedule the battery keeps absorbing reactive power during sags (B3, as low as −19.8 Mvar on S4 T1), and after long bolted faults the data-center load returns only when its 5 s cooling restore timer expires, at 5.00–5.56 s (L3), which misses the ERCOT load-return requirement. The saved ANDES result files for S1 T1 and S1 T2 are identical, so those two cases count as one independent result.

Fifteen of the 44 runs were repeated without the project. The authors selected these 15 after with-project slips or speed-gate stops; they form a targeted sample. In 13 of the 15 pairs, the same machines lose synchronism with and without the project. The other 29 runs have no without-project comparison. In S4 T8 (the 2,587 MW trip at the 115 kV tap), Buchanan unit 1 at bus 3105 reaches the 1.20 pu speed gate at 4.01 s with the project (1.229 pu), whereas without the project its speed stays between 0.985 and 1.007 pu for 60 s (Fig. 1). In S2 T11 (overvoltage), Buchanan slips only with the project, but the no-project twin stopped at 8.41 s, on another machine at bus 6215, so that attribution rests on a truncated comparison.

The rotor-angle result is RMS-only, because the machines that decide it lie outside the EMT area. Buchanan unit 1 (bus 3105) is eight branches from the POI, with a minimum series impedance of 0.418 pu on a 100 MVA base. The EMT area holds only the three Austin 2 turbines, and no EMT trace records rotor speed.

### B. *EMT Intake and System Study*

The EMT intake bench ran the plant on a single machine against an infinite bus at DPsim 10 and 20 µs. Of 220 rows, 176 pass, 2 fail (one voltage-step test at both time steps), and 42 are informational, so 178 rows are scored. The system study has 12 events: faults of several types next to the POI (E1–E4), a stuck breaker (E5), a 500 kV fault (E6), a six-sag replay (E7), sags (E8, E13), a line opening and angle step (E9), an overvoltage (E10), and a load trip (E11). E12 replays the RMS generation trip; the 84 rows cover the other events. A row is one dispatch, event, ride-through variant, counting state, and time step. The 42 cases (18 on S1, 14 on S2, and 5 each on S3 and S4), each run at 10 and 20 µs, give 84 rows. The S4 EMT runs used a plant output of +130 MW in place of the −133 MW charging dispatch specified for S4. Each row has ten lines, scored pass, fail, not applicable, or informational, giving 460 pass, 16 fail, 112 not applicable, and 252 informational lines. The failures include a no-return load variant, which fails the load-return line by design. The study also shows an overvoltage trip at fault clearing on S2 E4, which the RMS model cannot represent. Within DPsim, the two steps give the same scored verdicts on the 42 pairs, which is a within-tool check only.

All faults start at 0.5 s. E1 and E5 are bolted three-phase faults at 0.001 Ω, E2 is a single-line-to-ground fault on phase A for five cycles with 0.5 Ω, and E13 is a 5.5 Ω fault for 1.0 s on S1 only. DPsim applies a fault as a switched branch. ParaEMT applies a per-phase fault conductance (phase to ground for single- and double-line-to-ground faults, phase to phase for line-to-line faults) by matrix update and refactorization on the same schedule. After an LCL filter was added to the ParaEMT plant, the system study reports S2 E2 pre-fault values of 1.0391 pu and −131.1 MW in ParaEMT and 1.0394 pu and −131.4 MW in DPsim, with the plant in service and no locked line failed.

### C. Descriptive Cross-Tool Observations

Table III lists S1 values for two paired events (E1/T1 and E2/T5). They are unaligned observations. The tools differ in network, plant model, controls, operating point, and base (Table II), and the faults are represented differently. The prefault battery reactive power is 26.4 Mvar in ANDES, the data-center reactive schedule carried by the battery, and every EMT run uses a zero reactive order, which gives 1.4–1.6 Mvar at 24.9 kV. Reporting changes from each model's own initial condition leaves this difference in place.

TABLE III. OBSERVED S1 VALUES FOR PAIRED EVENTS (UNALIGNED)

| Observation | ANDES | DPsim | ParaEMT |
|---|---|---|---|
| Pre-fault battery P (MW) | 130.0 | 130.35 | 130.00 |
| Pre-fault Q (Mvar) | 26.4 | 1.6 | 1.4 |
| E1/T1 bolted, V at 6255 (pu) | 0.142 | 0.000 | 0.000 |
| E2/T5 SLG, V at 6255 (pu) | 0.725 | 0.713 | 0.820 |
| E2/T5 min-phase V (pu) | n/a | 0.149 | 0.109 |

EMT pre-fault P (mean over 0.30–0.48 s) and E2/T5 minimum V are from saved 50 µs traces sampled once per cycle; other EMT values are DPsim at 20 µs and ParaEMT at 50 µs.

In the bolted cases, the ANDES fault is a 0.001 pu shunt that holds 0.142 pu at the POI, whereas the EMT fault drives the POI voltage below 0.001 pu. For E2/T5, the DPsim and ParaEMT disturbances differ, with POI voltages of 0.713 and 0.820 pu and minimum-phase voltages of 0.149 and 0.109 pu, so that pair compares different disturbances.

## V. ERRORS AND INTERVENTIONS

Table IV lists the errors the authors identified in the agent logs and reports, classified by the authors from their inspection. Two were caught by the agents' own gate or cross-check (rows 1 and 2). One was self-reported (row 8), and one came from a review of the EMT report that an author communicated to the agent (row 9). Six were caught by author oversight (rows 3–7 and 10).

TABLE IV. ERRORS, DETECTION, AND RESOLUTION

| # | Error | Caught by | Resolution |
|---|---|---|---|
| 1 | Step-size gate failed twice | Agent gate | Located flag switching |
| 2 | Export had G and B swapped | Agent cross-check | Corrected export (+7.193 Mvar) |
| 3 | Six-sag runs stopped at 77.3 s | Author oversight | Half-step retries |
| 4 | RMS matrix too slow (Jacobian rebuilt) | Author oversight | Reuse and 1/240 s step |
| 5 | Reuse gave false collapse on light load | Author oversight | Fresh-matrix retry |
| 6 | Real 2.6 pu runaway treated as numerics | Author oversight | Reclassified as slip |
| 7 | 180° with 3% speed stop gave false stops | Author oversight | Speed gate and COI rule |
| 8 | ParaEMT light-load model invalid in 3 tries | Agent self-report | DPsim governs S2 |
| 9 | Post-clearing overvoltage up to 2.39 pu | Report review | Current chop fixed |
| 10 | First EMT report omitted intake bench | Author oversight | Bench added (220 rows) |

Speed was the most costly error (rows 4 and 5). Each Newton iteration of the RMS solver rebuilt and refactored the network Jacobian, so the case matrix advanced slowly. An author observed that an RMS run taking days indicated a fault, and the agent then profiled its solver. The fix combined Jacobian and factorization reuse with a step change from 1/480 to 1/240 s, so the gain is credited to the two changes together. The same reuse then produced a false voltage collapse on the light-load case, which advanced only 3–12 simulated seconds in two hours.

The step-size gate (G5) required POI voltage, battery P, and battery Q to agree within 0.01 pu at every sample, for 1/240 s against 1/480 s and for 1/480 s against 1/960 s. It failed twice (battery P differences of 0.147 pu at 1.096 s and 0.171 pu at 1.1126 s) and then passed, with maximum differences of 0.00387, 0.00646, and 0.00650 pu (voltage, P, Q) and 0.00193, 0.00195, and 0.00112 pu for the two comparisons. A later solver-reuse check against 1/480 s gave 0.0032, 0.0065, and 0.0064 pu on S1 T1 and within 0.0008 pu on S1 T5. Other cases were not compared against 1/480 s. For row 1, the REECA1 voltage-dip flag was changed to switch at its located threshold crossing, after which the gate passed on the third attempt.

In the EMT work, the overvoltage came from a breaker model that opened in one 50 µs step with no snubber, which chopped the fault current. A resistance ramp (a few ohms in the first millisecond, then exponential growth to 1 MΩ) lowered the S1 E5 post-clearing peak from 2.39 pu to 1.01 pu (minimum-phase one-cycle RMS voltage at the 230 kV POI within 0.5 s after clearing) and from 27.88 pu to 1.09 pu (highest instantaneous sample within 40 ms after clearing, lasting about 3–8 ms; recorded in the tuning log only). These pairs come from an earlier plant configuration in DPsim at 50 µs; the ramp was carried into the 84-row study at 10 and 20 µs. The physical validity of the ramp was not tested.

Authors sent 63 recorded messages: 55 to the RMS agent (38 check-ins and 17 other messages) and eight to the EMT agent. The elapsed time from the first author message (Oct. 29, 3:44 PM MST) to the integrated Interconnection Study Report (Nov. 2, 10:27 AM MST) was 3.8 days. The EMT log ends before its intake and system-study phase finished, so the EMT count and the intervention total are lower bounds.

## VI. Limitations

The study supports observations about workflow execution. The RMS and EMT models differ in network, plant model, controls, operating point, base, and initial conditions, so Table III places each tool's values side by side for reference. The Kron equivalent is identified by its recorded hash, and its parameter file was not kept in the archived packages; it was checked against the full network for short-circuit level and power flow only. The reduced network is phase decoupled with Z0 = Z1, so unbalanced EMT faults use the positive-sequence impedance for the zero-sequence path. The REECA1 flag switching and the EMT breaker ramp remain to be validated against physical behavior. Saved EMT traces, at one sample per cycle, cover 38 DPsim and 25 ParaEMT runs of the 84 scored rows, with S2 traces from DPsim only. The applied loss-of-synchronism rule differs from the locked criterion text, and the authors both wrote the criteria and reviewed the results.

A validated RMS and EMT comparison would need a shared operating point, a documented and validated network equivalent, and an independent or industry reference.

## VII. Conclusion

Two AI agents executed a predefined RMS and EMT ride-through workflow on a 2,000-bus synthetic Texas grid with open-source tools. Across the 44 RMS cases, 127 criterion cells fail, including loss of synchronism in all 44, and two findings concern the project plant (B3 and L3). In 13 of the 15 selected without-project comparisons the same machines slip without the project. The agents' own checks caught two of the errors in Table IV, and the authors caught six. Modeling judgment and the treatment of failed runs required an engineer.